\documentclass[a4paper,12pt]{revtex4}
\usepackage{physics}
\usepackage{graphics}
\usepackage{color}
\usepackage[dvips]{graphicx}

\usepackage{pdfpages}
\usepackage{epstopdf}

\begin{document}

\title{Coherent population trapping in a Raman atom interferometer}


\author{B. Cheng, P. Gillot, S. Merlet, F. Pereira Dos Santos}                  

\affiliation{\\
LNE-SYRTE, Observatoire de Paris, PSL Research University, CNRS, Sorbonne Universit\'{e}s, UPMC Univ. Paris 06, 61 avenue de l'Observatoire, 75014 Paris, France}

\begin{abstract}

We investigate the effect of coherent population trapping (CPT) in an atom interferometer gravimeter based on the use of stimulated Raman transitions. We find that CPT leads to significant phase shifts, of order of a few mrad, which may compromise the accuracy of inertial measurements. We show that this effect is rejected by the k-reversal technique, which consists in averaging inertial measurements performed with two opposite orientations of the Raman wavevector k, provided that internal states at the input of the interferometer are kept identical for both configurations.
\end{abstract}

\maketitle

\section{Introduction}

Gravimeters based on Mach Zehnder type atom interferometer reach nowadays long term stabilities in the low $10^{-10}g$ range \cite{Gillot2014,Hu2013} or better \cite{Freier2015} and accuracies of a few $10^{-9}g$ \cite{Peters2001,Francis2015,Freier2015}, comparable to classical corner cube gravimeters \cite{Niebauer1995}. On going efforts to improve the stability of cold atom gravimeters focus on strategies to accurately determine or reject interferometer phase fluctuations arising from changes of the experimental parameters (such as due to light shifts and Doppler shifts fluctuations \cite{Gillot2016}) or from environmental effects (via for instance the direct comparison of two gravimeters, eventually based on different technologies \cite{Freier2015}).

A common and very efficient method consists in alternating the direction of the Raman wavevector, which allows rejecting the phase shifts which are independent of the Raman laser wavevector direction. This rejection is in practice limited by the difference of the trajectories of the atoms between these two interferometer configurations, due to the change in the direction of the momentum kick imparted to the atoms by the lasers. To be quantitative, the maximum position shift between these trajectories reaches, for our total interferometer duration of 160 ms, up to $2~$mm in the vertical direction.

It is thus of interest to find methods that maximize the trajectories overlap when changing the direction of the Raman wevector. As already pointed out in \cite{Mehlstaubler2007}, this can be realized for instance by changing the internal state of the atom at the input of the interferometer. The momentum kick then occurs in the same direction, despite the change of the direction of the Raman wavevector. We show here that this technique has a drawback, and leads to a bias in the measurement of gravity, arising from a phase shift linked to coherent population trapping (CPT).  
The effect of CPT was put into evidence in \cite{Butts2011} by measuring dark-state coherences and population differences induced in cold cesium atoms by velocity-sensitive and velocity-insensitive Raman pulses. It was also claimed in \cite{Butts2011} that CPT effects should lead to spurious phase shifts of order of a few mrad in Mach Zehnder interferometer, which the measurements we present here confirm. 

In this article, we perform a detailed evaluation of the phase shift induced by CPT effects. We first investigate this effect theoretically following the formalism developed in \cite{Butts2011} and extending it to the case of a Raman interferometer. We show results of measurements where we exchange internal states at the input of the interferometer to put this effect in evidence. We study in particular its dependence on relevant parameters of the Raman laser, such as one-photon Raman laser detuning, Raman pulses and interferometer duration.

\section{Theory}

We measure gravity using an atom interferometer realized by counterpropagating Raman transitions. Raman transitions are two-photon transitions which couple two states $\ket{g}$ and $\ket{e}$ (in our case two hyperfine ground states of an alkali atom) via the off-resonant excitation of an excited state $\ket{i}$. CPT effects arise from the dynamics of this 3 level system ($\ket{g}$, $\ket{e}$, $\ket{i}$) interacting with the Raman lasers, when taking into account the influence of spontaneous emission from the excited level. In \cite{Butts2011}, the evolution of a three level system in the field of two lasers is developed in the interaction picture taking into account spontaneous emission. The density matrix $R_{int}$ of the three states is given by:
\begin{equation} 
\frac{\dd{R_{int}}}{\dd{t}}= [\frac{1}{i\hbar} (\hat{V}_{int}-\hat{H}_{int}), R_{int} ]+R_{SE}
\end{equation}
where $ \hat{H}_{int}$ is the laser energy, $\hat{V}_{int}$ is the coupling in the interaction picture and $R_{SE}$ is the spontaneous decay of the density matrix.

Adiabatic elimination of the excited state $\ket{i}$ allows to derive differential equations governing the dynamics of the system in the basis restricted to the two states $\ket{g}$ and $\ket{e}$ \cite{Butts2011}. These are given by eq. \ref{eq:eq2}, where $\Gamma$ is the linewidth of the excited state and $\Omega_{eff}$ is the effective 2-photon Rabi frequency. $\delta (t)-\delta _{\text{AC}}$ is the two-photon Raman detuning, $\Delta$ is the one-photon Raman laser detuning from the excited state and $\delta_{AC}$ is (the one-photon) differential light shift.

\begin{eqnarray}
\text{$\rho _{ee}$}'(t)+\Im(\text{$\Omega _{eff}$} \text{r$_{eg}$}(t))+\frac{\Gamma  \Re(\text{$\Omega _{eff}$} \text{r$_{eg}$}(t))}{2 \Delta }+\frac{\Gamma  \Omega _{\text{eAC}} \text{$\rho _{ee}$}(t)}{\Delta }=0 \nonumber\\
\text{$\rho _{gg}$}'(t)-\Im(\text{$\Omega _{eff}$} \text{r$_{eg}$}(t))+\frac{\Gamma  \Re(\text{$\Omega _{eff}$} \text{r$_{eg}$}(t))}{2 \Delta }+\frac{\Gamma  \Omega _{\text{gAC}} \text{$\rho _{gg}$}(t)}{\Delta }=0 \nonumber\\
\text{r$_{eg}$}'(t)-\frac{1}{2} i \text{$\Omega _{eff}$}^* (\text{$\rho _{ee}$}(t)-\text{$\rho _{gg}$}(t))+\frac{\Gamma  \left(\Omega _{\text{eAC}}+\Omega _{\text{gAC}}\right) \text{r$_{eg}$}(t)}{2 \Delta }\nonumber\\
-\text{i r$_{eg}$}(t) \left(\delta (t)-\delta _{\text{AC}}\right)+\frac{\Gamma  \text{$\Omega _{eff}$}^* (\text{$\rho _{ee}$}(t)+\text{$\rho _{gg}$}(t))}{4 \Delta }=0
\label{eq:eq2}
\end{eqnarray}

\begin{figure}
	\centering
	\includegraphics[width=14 cm]{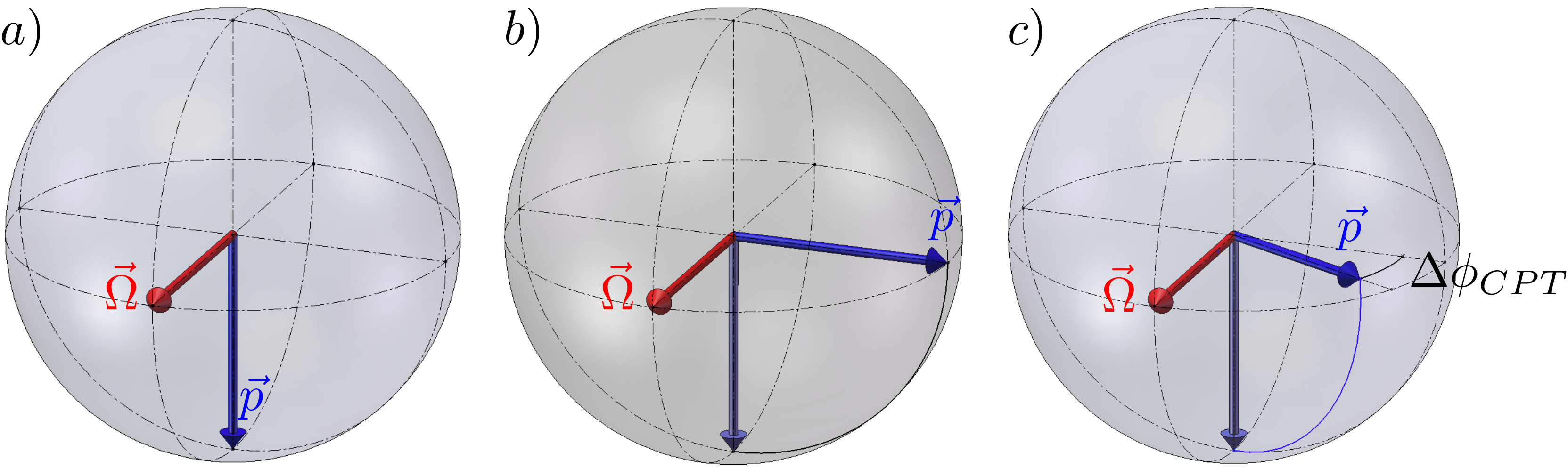}
	\caption{The evolution of the pseudo-spin during a $\pi/2$ Raman pulse in the Bloch sphere with and without CPT effects. a) represents the initial vector state pointing down and the Raman drive $\Omega$, b) the ideal situation of a perfect $\pi/2$ pulse without spontaneous emission, and c) the case with spontaneous emission.}
	\label{bloch}
\end{figure}

A detailed analysis of the evolution of the system is done in \cite{Butts2011}, where spontaneous emission is shown to lead to coherent population trapping. For on resonance driving, the system asymptotically evolves towards a dark state, uncoupled to the Raman lasers. Representing the quantum state as a vector in the Bloch sphere helps understanding the phase shift introduced by the CPT effect in our situation, where the duration of Raman pulses is more than two orders of magnitude shorter than the characteristic time of evolution into the dark state. In this picture, the atomic state is depicted by the pseudo spin ($\vec{P}$). While ($\vec{P}$) rotates in a plane perpendicular to the Raman vector ($\vec{\Omega}$) during the Raman pulse, spontaneous emission makes the pseudo-spin move off this plane. For short Raman pulse ($\frac{\Gamma\Omega_{eff}}{2\Delta}\tau\ll 1$), this off-the-plane shift increases linearly with time, at a rate $\frac{\Gamma\Omega_{eff}}{2\Delta}$, independent of the one photon transition couplings ($\Omega _{\text{gAC}}$, $\Omega _{\text{eAC}} $).
This dynamic is illustrated for a $\pi/2$ Raman pulse in fig. \ref{bloch}. Starting from a initial state pointing downward in the Bloch sphere (displayed in a)), the drive ($\vec{\Omega}$) induces in the absence of spontaneous emission a rotation of the vector state by $\pi/2$ in the plane perpendicular to the direction of the drive. The final state then lies, as displayed in b), in the equatorial plane, perpendicular to the drive. Taking into account spontaneous emission, we find that the final pseudo-spin is reduced in amplitude and shifted by an angle $\Delta \phi_{CPT}$ in the equatorial plane, as displayed in c).

The CPT phase at resonance (we do not consider any detuning from the Raman resonance condition here) is found to be approximately given by: 
\begin{equation}
\Delta\phi_{CPT}=\frac{\Gamma\tau\Omega_{eff}}{2\Delta}
\label{eq:cpt1}
\end{equation}
where $\tau$ is the pulse duration of the Raman pulse. 

To evaluate the amplitude of the effect, we consider the case of $^{87}$Rb atoms, with Raman lasers at one-photon Raman detuning $\Delta~=~-0.932$ GHz, and for a Raman pulse duration corresponding to a $\pi/2$ pulse. We calculate a phase shift of 5.06 mrad, which is significant when seeking for precise gravity measurement. 

This CPT phase leads to an interferometer phase shift in three-pulse interferometers based on Raman transitions, which arises from the effect of the first pulse only, as already claimed in \cite{Butts2011}. Indeed, the second and third pulses, though they contribute to increase the population of the dark state, do not lead to additional phase shifts. The second pulse adds the same phase shift to both interferometer arms, while the third pulse creates a polarisation in the equatorial plane which does not affect the final state population. 

For a comparison of the CPT induced phase shift with measurements in a real interferometer, detunings due to the Doppler effect need to be considered. 
For that purpose, we performed a numerical evaluation of the interferometer phase shift by numerically solving the equations of evolution of the density matrix for the three pulse sequence and averaging the calculated transition probability of the interferometer over the Doppler distribution (linked to the velocity distribution). In order to simulate the interferometer fringe pattern, we repeat the calculation for increasing values of a controlled phase offset applied at the third pulse to the Raman lasers. From a fit of the fringe pattern, we finally extract the CPT induced phase. We calculated with this simulation the phase shift for the interferometer parameters given above. The Rabi frequency is chosen to be 11.4 kHz. The pulses durations are $22-44-22~\mu$s, which correspond to a $\pi/2-\pi-\pi/2$ pulse sequence. The initial velocity distribution is taken to be Gaussian, with $\sigma_v \sim 2\hbar k_L/m_{Rb}$, where $m_{Rb}$ is the mass of a $^{87}$Rb atom and $k_L$ is the photon momentum at 780 nm. In addition, we consider that the atoms are velocity selected with a Raman $\pi$ pulse of duration $44~\mu$s before entering the interferometer (as we will do later in the experiment). We find for these parameters a phase shift $\Delta\phi$ of 5.35 mrad. This differs from the result of eq. \ref{eq:cpt1} by about 6\% only, which indicates that the average over the velocity distribution has a limited influence on the result. Moreover, with the simulation, we confirm that the effect on the interferometer phase is given by the CPT phase of the first pulse. Finally, the calculated phase shift corresponds to a bias on the g measurement of $\Delta g~=~\Delta\phi/kT^2 =~5.2~\mu$Gal , where $k \simeq 2k_L$ is the effective Raman wavevector, and 1 Gal~=~1 cm/s$^2$. 

Hopefully, this phase shift is independent of the Raman wavevector direction. It is thus in principle well rejected by the k-reversal technique, which consists in averaging the measurements performed using two opposite directions of the Raman effective wavevector $k$. Yet, as a remarkable feature, we find that this phase shift changes sign when the internal state at the input of the interferometer is changed. For the k-reversal rejection to hold, it is thus mandatory that the internal state at the input of the interferometer is the same for both directions of $k$.

\section{Experiments}

To put the CPT effect into evidence and evaluate its influence, we exploit its dependence on the internal state at the input of the interferometer. We will thus perform differential measurements of the gravity acceleration $g$ for given directions of the Raman wavevector, but with different internal states of the atom at the input of the interferometer. 

The experimental setup is described in detail in \cite{Louchet-Chauvet2011}. We briefly recall here the main phases of the experimental sequence. We start by trapping a few $10^7$ atoms in a 3D Magneto-Optical Trap for 80~ms. A subsequent molasses phase cools the atoms down to a temperature of 2~$\mu$K. The molasses beams are then switched off and the atomic cloud is let to fall. After a preparation phase detailed below, we drive a three pulse Mach-Zehnder type Raman interferometer, with a total interferometer time of $2T=160$ ms, where $T$ is the separation time between consecutive pulses. The populations in the two output ports of the interferometer are finally measured via a state selective fluorescence detection setup at the bottom of the vacuum chamber. 

For the  preparation of the atomic state at the input of the interferometer, we normally apply 2 microwave pulses. The first one is used for the sub-$m_F$ state selection into the state $|F=1,m_F=0\rangle$. It transfers atoms in the $|F=2,m_F=0\rangle$ into the $|F=1,m_F=0\rangle$, and is followed by a pulse of a pusher beam that removes atoms remaining in the $|F=2\rangle$ state.
The second one is used to retransfer the atoms into the $|F=1\rangle$ internal state before the velocity selection occurs. This selection is realized with a Raman pulse (that transfers the centre of the velocity distribution back into the state $|F=1,m_F=0\rangle$) and a subsequent second pulse of the pusher beam. The use of a second microwave pulse is required as we do not have a pusher beam resonant with $|F=1\rangle \rightarrow |F'\rangle$ transition. The final internal state at the input of the interferometer is thus $|F=1, m_F=0\rangle$. 
To prepare the atoms into the $|F=2, m_F=0\rangle$ state at the input of the interferometer, a possibility would be to simply apply a third microwave pulse after the normal sequence. In this way, though, the velocity kicks imparted by the selection and Raman pulses would occur in the same direction, which would modify the trajectories of the interferometer paths. As an alternative, we remove the second microwave pulse, so that the velocity selection is performed from $|F=1\rangle$ to $|F=2\rangle$. Then, we get rid of the atoms that are not velocity selected with a sequence comprised of two microwave $\pi$ pulses and a pulse of pusher beam in between them.

\begin{figure}[h]
	\includegraphics[width=12 cm]{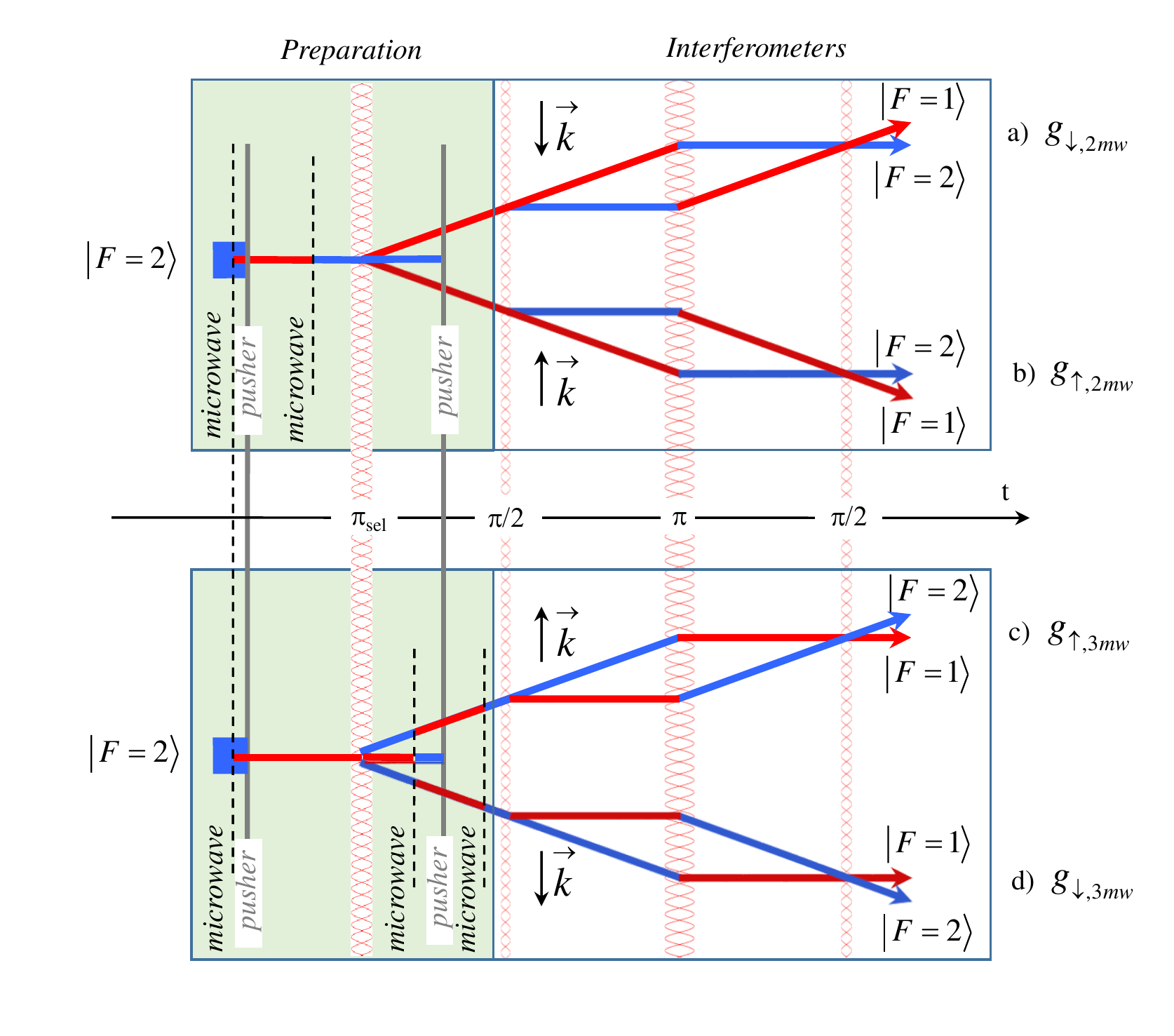}
	\caption{Different preparation sequences, with two or three microwave pulses, corresponding to input states in $|F=1\rangle$ or $|F=2\rangle$. The corresponding interferometer configurations with the trajectories along the two interferometer paths are also displayed.}
	\label{fig:statePrepare}
\end{figure}

The different preparation sequences and the corresponding interferometer configurations we use for the gravity measurements performed here are shown in figure~\ref{fig:statePrepare}. 

Usually, we use two interleaved measurements with opposite wavevectors (displayed as case a) and b) in figure~\ref{fig:statePrepare}) with atoms entering the interferometer in the state $|F=1\rangle$, which requires two microwave pulses in the preparation. The gravity measurement is then obtained from the average of the two measurements.
Case c) and d) correspond to a different preparation sequence, using three microwave pulses, with atoms entering the interferometer in the state $|F=2\rangle$. One can note that the trajectories of the atomic wavepackets along the two interferometer paths are the same for the $k_{\downarrow}$ interferometer using two microwave pulses (case a)) and the $k_{\uparrow}$ interferometer using three microwave pulse (case c)). The same holds for the $k_{\uparrow}$ interferometer using two microwave pulses (case b)) and the $k_{\downarrow}$ interferometer using three microwave pulses (case d)).
This allows to realize interleaved measurements with $k_{\uparrow}$ and $k_{\downarrow}$ interferometers while keeping the trajectories overlapped. It simply requires to replace for instance the $k_{\uparrow}$ interferometer of case b) by the $k_{\uparrow}$ interferometer of case c) (or the $k_{\downarrow}$ interferometer of case a) by the $k_{\downarrow}$ interferometer of case d)). 

We show now that the change of internal state at the input of the interferometer which is associated with this swap makes the new pair of configurations sensitive to CPT effect. We present in the following measurements of the difference in the phases (and the corresponding differences in the measured values of g) between the $k_{\uparrow}$ interferometers of case b) and c), and the difference between the $k_{\downarrow}$ interferometers of case a) and d).

\begin{figure}[h]
	\includegraphics[width=3 in]{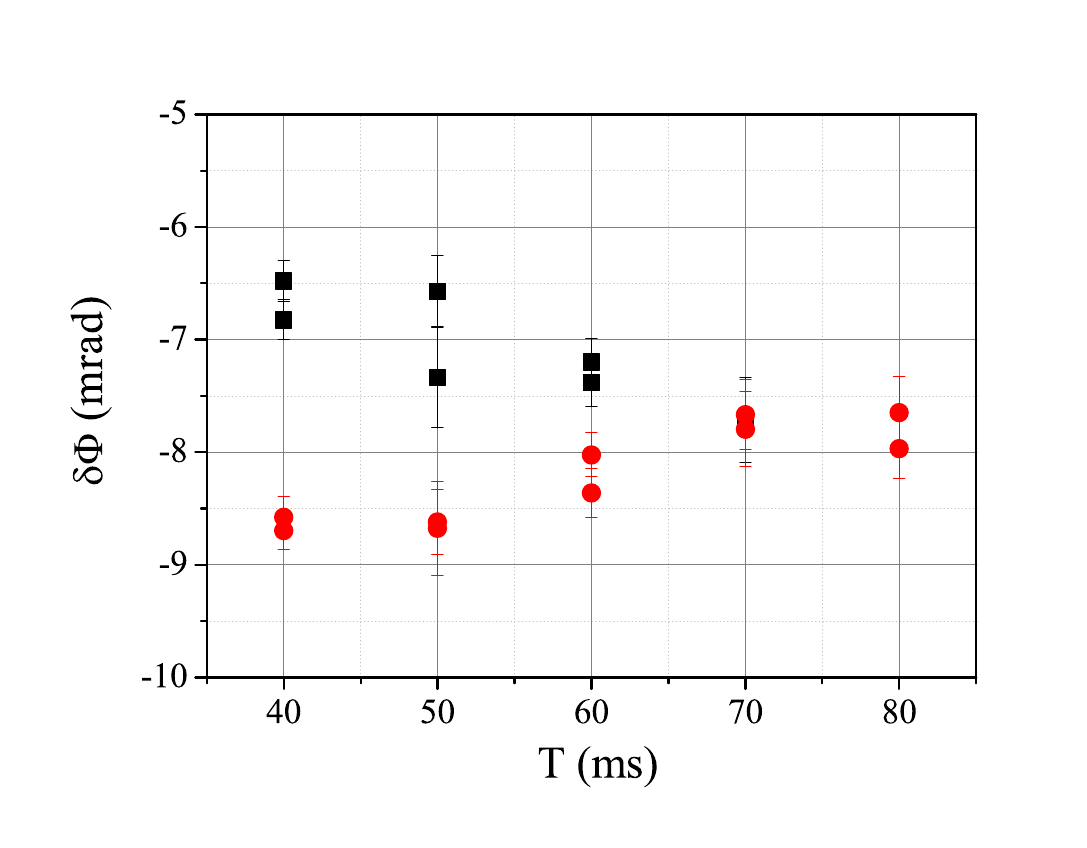}
	\includegraphics[width=3 in]{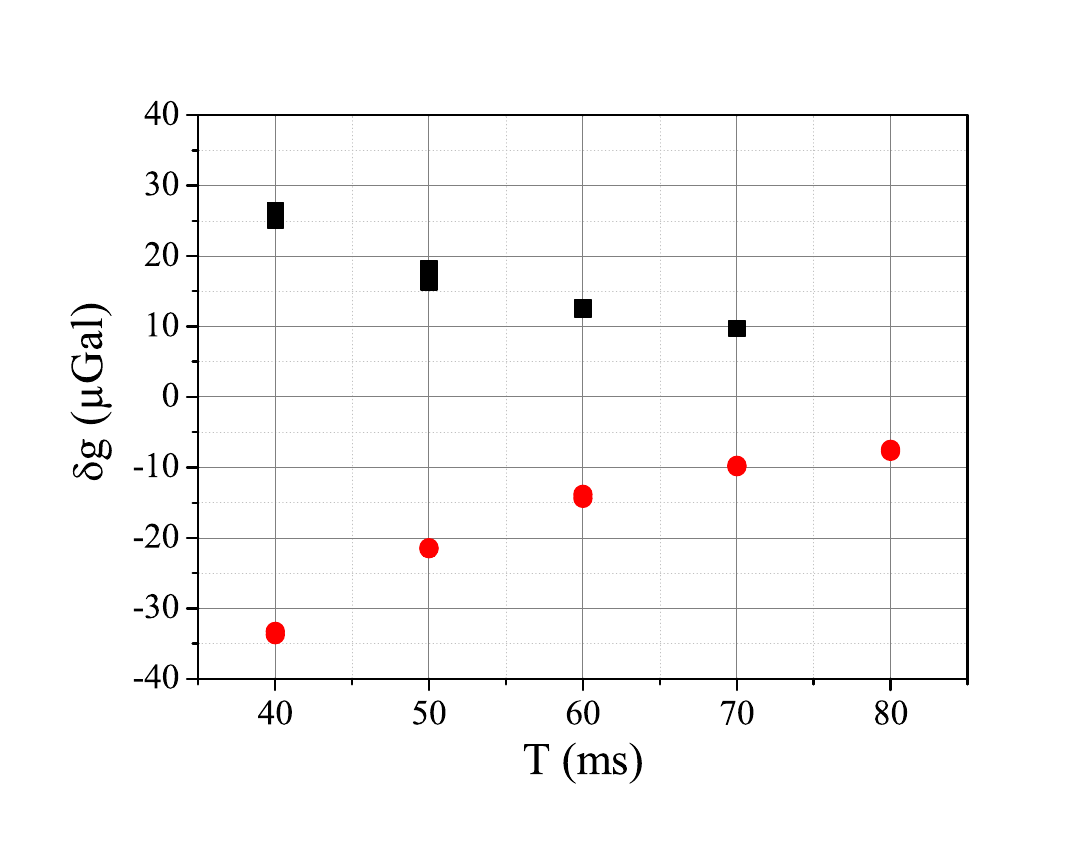}
	\caption{Differences in the interferometer phases and in the measured g values for input states in different hyperfine states as a function of $T$, ranging from 40 to 80 ms. Black squares: $k_{\downarrow}$ interferometers, Red circles: $k_{\uparrow}$ interferometers. The one-photon detuning of the Raman lasers is -0.9 GHz.}
	\label{fig:T}
\end{figure}

Figure~\ref{fig:T} displays the measured differences in the interferometer phases as a function of the Raman pulse spacing $T$. We find small variations with $T$ of these differences, with opposite trends for $k_{\uparrow}$ and $k_{\downarrow}$ interferometers, which are not reproduced by the simple model above. We find on average a value of about 7.7(4) mrad in absolute value. As the CPT phase changes sign with the internal state, the measured difference in the interferometer phases is twice this CPT phase. We would thus expect differences of 10.7 mrad, which is significantly larger than our measurement. This difference may be explained by the fact that the our model neglects the detailed structure of the energy levels of the atoms (hyperfine structure of the excited state $i$, Zeeman sublevels ...).   
The interferometer phase difference corresponds to a difference in the g value of $7.7(4)~\mu$Gal for an interferometer duration of $2T~=~80~$ms. As the gravity phase shift scales as $T^2$, we find, as displayed in Figure~\ref{fig:T}, that the lower the separation time $T$, the higher the effect on the gravity value.

\begin{figure}[h]
	\includegraphics[width=4 in]{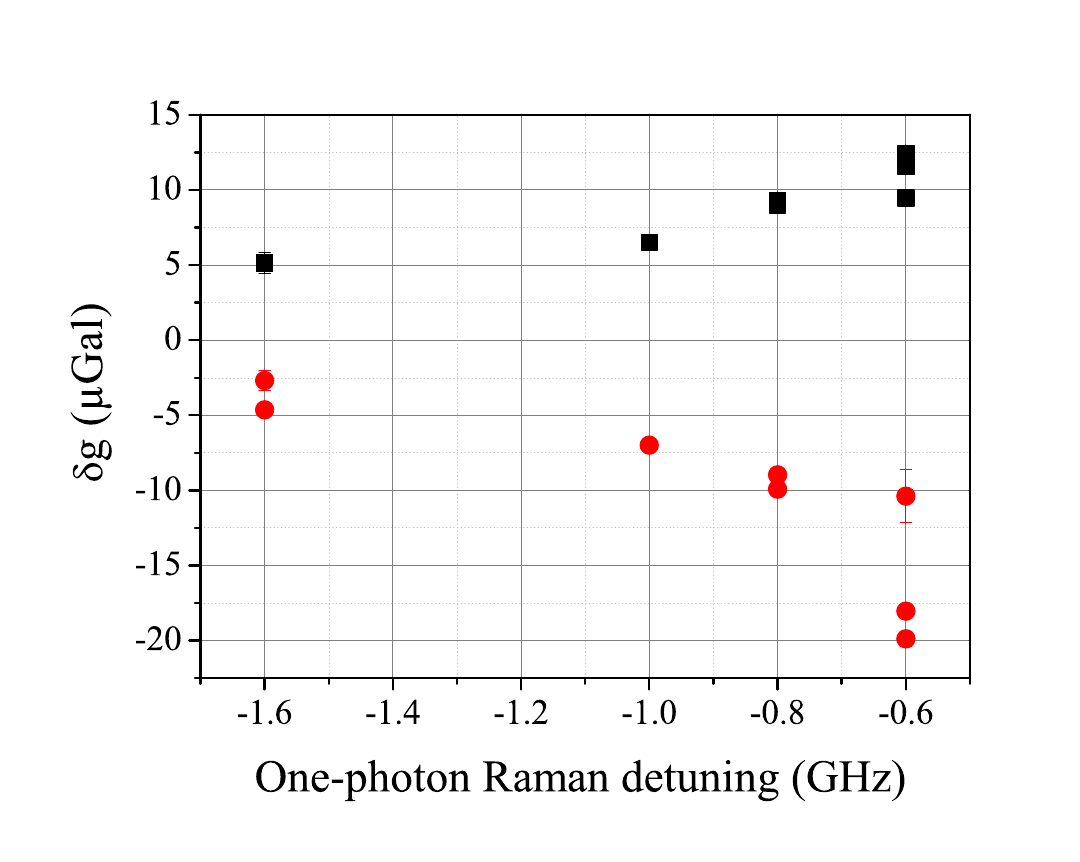}
	\caption{Differences in the measured g values for input states in different hyperfine states as a function of the one-photon laser detuning, ranging from -0.6 GHz to -1.6 GHz, for $T~=~80$ms. Black squares: $k_{\downarrow}$, Red circles: $k_{\uparrow}$.}
	\label{fig:Delta}
\end{figure}

We then measured the dependence of the phase shift with the one-photon laser detuning from the excited state $\Delta$, keeping the Rabi frequency constant, by adjusting the Raman laser intensity. The results, displayed on fig. \ref{fig:Delta}, confirm the expected scaling: the phase shift decreases inversely proportionally to $\Delta$ (see eq. \ref{eq:cpt1}), which we take as a strong evidence that the measured shift originates indeed from the effect of spontaneous emission.

\begin{figure}[h]
	\includegraphics[width=4 in]{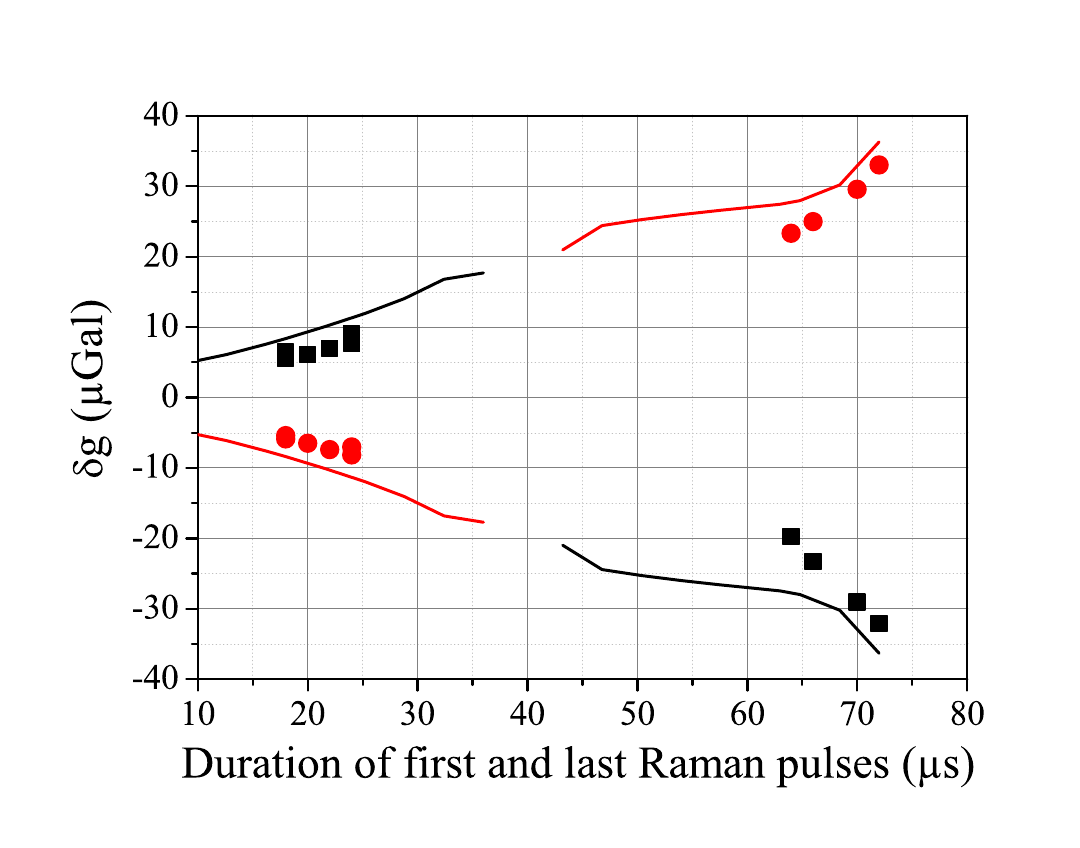}
	\caption{Differences in the measured g values for input states in different hyperfine states as a function of the duration of the first and third Raman pulses, for a Rabi frequency of $2\pi\times 11.4$ kHz and $T=80$ ms. The duration of the second pulse is kept constant at $44~\mu$s. Black squares: $k_{\downarrow}$, Red circles: $k_{\uparrow}$. Lines: calculations.}
	\label{fig:tau}
\end{figure}

Finally, we measured the variation of the CPT induced phase shift with the duration of the first Raman pulse, for a fixed Rabi frequency of $2\pi\times 11.4$ kHz, and compared these measurements with the results of the numerical simulations. We performed measurements for values ranging from 17 to 24 $\mu$s (close to the duration of 22 $\mu$s of the perfect $\pi/2$ pulse) and 64 to 72 $\mu$s (close to a $3\pi/2$ pulse). The results are displayed on fig.~\ref{fig:tau}. The shift on the measurement of g increases with increasing durations, and changes sign when the pulse becomes longer than a $\pi$ pulse. The trends we measure are in good agreement with the results of the numerical simulation, which are displayed as lines, though the quantitative agreement is here again not perfect.

\section{Conclusion}

We have studied the effect of CPT in an atom gravimeter, based on an Mach-Zehnder type atom interferometer, realized with a sequence of three Raman pulses. Measurements of the phase shift induced by this effect, and thus of the corresponding bias onto the measurement of gravity, have been performed as a function of the parameters of the Raman lasers and of the pulse sequence, such as pulse duration, and detuning of the Raman lasers. The trends in the measurements are found to be in good agreement with the behaviour derived from calculations based on a simple three level model. A better match between measured and calculated phase shifts would certainly require a model which takes into account the real internal structure of the atom and the polarization state of the Raman lasers.

This phase shift is a drawback when alternating interferometer measurements with configurations that change not only the direction of the Raman wavevector but also the internal state at the input of the interferometer. Indeed, it  changes sign with configuration, as does the gravity phase shift. This finally results in a bias in the determination of g, when averaging the g measurements over the two configurations. However, changing the internal state at the input of the interferometer offers a better superposition of the trajectories between these two configurations. This allows for a better rejection of magnetic field gradients \cite{Mehlstaubler2007} and eventual light shift longitudinal inhomogeneities. In that case, though, the measured g value needs to be corrected for the phase shift induced by CPT effects.   

\section{Acknowlegments}

B. C. thanks the Labex First-TF for financial support.

\end{document}